\documentclass[times,twocolumn]{aastex6}

\usepackage{graphicx}

\usepackage{enumitem}
\usepackage{amsmath}
\usepackage[perpage]{footmisc}
\usepackage{multirow}
\usepackage{booktabs}
\usepackage{array}
\usepackage{dcolumn}
\usepackage{tablefootnote}

\usepackage{CJK}

\usepackage{etoolbox}
\makeatletter
\patchcmd{\NAT@citex}
  {\@citea\NAT@hyper@{%
     \NAT@nmfmt{\NAT@nm}%
     \hyper@natlinkbreak{\NAT@aysep\NAT@spacechar}{\@citeb\@extra@b@citeb}%
     \NAT@date}}
  {\@citea\NAT@nmfmt{\NAT@nm}%
   \NAT@aysep\NAT@spacechar\NAT@hyper@{\NAT@date}}{}{}
\patchcmd{\NAT@citex}
  {\@citea\NAT@hyper@{%
     \NAT@nmfmt{\NAT@nm}%
     \hyper@natlinkbreak{\NAT@spacechar\NAT@@open\if*#1*\else#1\NAT@spacechar\fi}%
       {\@citeb\@extra@b@citeb}%
     \NAT@date}}
  {\@citea\NAT@nmfmt{\NAT@nm}%
   \NAT@spacechar\NAT@@open\if*#1*\else#1\NAT@spacechar\fi\NAT@hyper@{\NAT@date}}
  {}{}
\makeatother

\shorttitle{High-Cadence Vector Magnetograms}
\shortauthors{Sun et al.}

\begin{document}

\begin{CJK}{UTF8}{}

\title{Investigating the Magnetic Imprints of Major Solar Eruptions\\
with \textit{SDO}/HMI High-Cadence Vector Magnetograms}

\author{
\begin{CJK}{UTF8}{gbsn} 
Xudong Sun (孙旭东)\altaffilmark{1}, J. Todd Hoeksema\altaffilmark{1}, Yang Liu (刘扬)\altaffilmark{1}, Maria Kazachenko\altaffilmark{2}, and Ruizhu Chen (陈瑞竹)\altaffilmark{1,3} 
\end{CJK}
}

\affil{
$^1$W. W. Hansen Experimental Physics Laboratory, Stanford University, Stanford, CA 94305, USA; \href{mailto:xudong@sun.stanford.edu}{xudong@sun.stanford.edu}\\
$^2$Space Sciences Laboratory, University of California, Berkeley, CA 94720-7450, USA\\
$^3$Department of Physics, Stanford University, Stanford, CA 94305, USA
}


\begin{abstract}
The solar active region photospheric magnetic field evolves rapidly during major eruptive events, suggesting appreciable feedback from the corona. Previous studies of these ``magnetic imprints'' are mostly based on line-of-sight only or lower-cadence vector observations; a temporally resolved depiction of the vector field evolution is hitherto lacking. Here, we introduce the high-cadence (90~s or 135~s) vector magnetogram dataset from the Helioseismic and Magnetic Imager (HMI), which is well suited for investigating the phenomenon. These observations allow quantitative characterization of the permanent, step-like changes that are most pronounced in the horizontal field component ($B_h$). A highly structured pattern emerges from analysis of an archetypical event, \texttt{SOL2011-02-15T01:56}, where $B_h$ near the main polarity inversion line increases significantly during the earlier phase of the associated flare with a time scale of several minutes, while $B_h$ in the periphery decreases at later times with smaller magnitudes and a slightly longer time scale. The dataset also allows effective identification of the ``magnetic transient'' artifact, where enhanced flare emission alters the Stokes profiles and the inferred magnetic field becomes unreliable. Our results provide insights on the momentum processes in solar eruptions. The dataset may also be useful to the study of sunquakes and data-driven modeling of the corona.
\end{abstract}


\section{Introduction}
\label{sec:intro}

Solar active regions (ARs) harbor strong magnetic fields that often carry significant electric currents. Processes such as flux emergence and shearing motion gradually bring excess magnetic energy into the low corona. During an eruption, the coronal magnetic field reorganizes rapidly, converting part of the magnetic energy into intense emission as flares, or propelling plasma into interplanetary space as coronal mass ejections (CMEs). There are two distinctive time scales in this ``storage and release'' picture \citep[e.g.][]{schrijver2009}. In the plasma-dominated photosphere, the characteristic Alfv\'{e}n speed ($v_A$) is low. Magnetic evolution leading to an eruption occurs over hours or days. In the lower corona, however, plasma $\beta$ is low and $v_A$ can reach a thousand kilometers per second. Flare emission and CME acceleration occur on a shorter time scale, on the order of 10 minutes.

Such a separation of time scales breaks down during major solar eruptions. There has been mounting evidence for the rapid evolution of the photospheric magnetic field associated with intense flares and fast CMEs \citep[for a recent review, see][]{wanghm2015}. For example, permanent and step-wise changes have been observed in the line-of-sight (LoS) field component ($B_l$) for many large flares \citep[e.g.,][]{cameron1999,kosovichev2001,sudol2005,petrie2010}. Changes up to several hundred Gauss occur within mere minutes. In general, the LoS magnetic flux on the disk-ward side of the AR decreases, while the limb-ward flux increases, indicating a more horizontal magnetic configuration near the polarity inversion line \citep[PIL;][]{wanghm2002,wanghm2010}. The pattern is consistent with the observed darkening of the inner penumbrae and weakening of the outer penumbrae in $\delta$-sunspots \citep{liuc2005}. The step-wise changes of $B_l$ often start in the early phase of a flare, well before the soft X-ray (SXR) peak \citep{cliver2012,johnstone2012,burtseva2015}.


\begin{figure*}[th!]
\centerline{\includegraphics{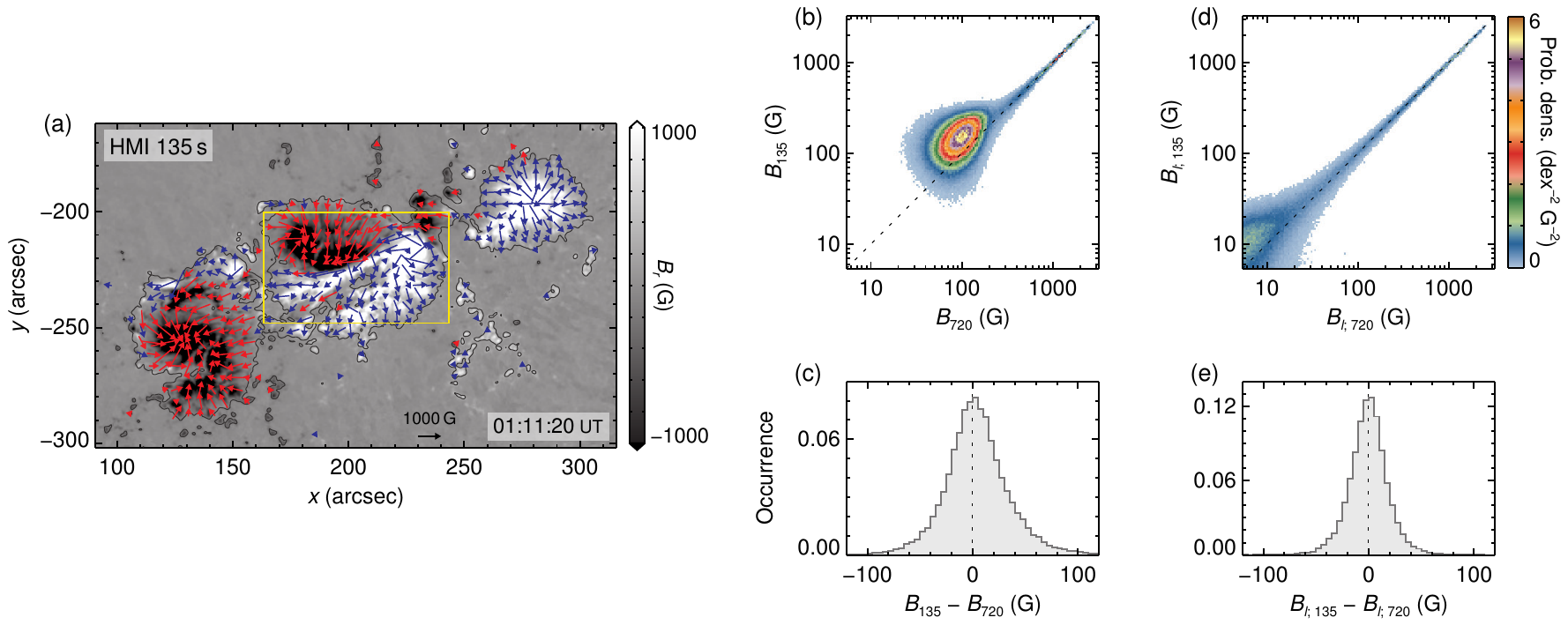}}
\caption{Comparison of 135~s and 720~s magnetograms for AR 11158 at 2011-02-15T01:11:20 UT. (a) Background image shows the 135~s radial field $B_r$. Arrows show horizontal field $B_h$. Contours show total field strength $B$ at 300~G. The yellow box outlines the field of view for Figures~\ref{f:profile}--\ref{f:stats}. (b) Two-dimensional distribution of $B$ for 135~s ($B_{135}$) and 720~s ($B_{720}$) data. Logarithm scale highlights the low field values. Dotted line has a slope of unity. (c) Distribution of $B_{135} -  B_{720}$ for pixels with $B_{135}, B_{720}>300$~G. (d) Similar to (b), but using only the absolute values of $B_l$ when the two versions agree in sign. (e) Similar to (c), but for $B_l$.}
\label{f:compare}
\end{figure*}


This picture is consistent with vector field observations, which showed that the horizontal field component ($B_h$) and the shear angle near the main PIL increases after a flare \citep{wanghm1992,wanghm1994}. In addition, $B_h$ has been found to decrease in the peripheral areas of $\delta$-sunspots \citep{wangjx2009}. Since 2010, routine full-disk vector magnetograms from the Helioseismic and Magnetic Imager \citep[HMI;][]{schou2012,hoeksema2014} aboard the \textit{Solar Dynamics Observatory} (\textit{SDO}) have provided definitive evidence that the rapid, permanent photospheric field changes occur during most large flares \citep[e.g.,][]{wangs2012a,wangs2012b,sun2012,petrie2012,petrie2013}. A common scenario is that $B_h$ increases significantly near the PIL, whereas the radial field component ($B_r$) varies less and without a clear pattern. The field becomes stronger and more inclined in the AR core.

The rapid appearance of these ``magnetic imprints'' suggests that they have a coronal origin, possibly as feedback from the eruption. In the ``coronal implosion'' conjecture \citep{hudson2000}, the non-erupting AR loops must contract to compensate for the loss of magnetic energy, which is consistent with the observed increase of field inclination. The impulsive coronal Lorentz force, which accelerates CME plasma to high speed, must act downward on the rest of the Sun due to momentum conservation \citep{hudson2008,fisher2012}. This back reaction has been evoked to explain sunquakes \citep{zharkov2011,alvaradogomez2012} and sudden changes of sunspot rotation rate during flares \citep{wangs2014,liuc2016}. Flares without a CME seem to exhibit weaker magnetic imprints than its eruptive counterpart \citep{sun2015}.

Past observations have revealed much about the nature of magnetic imprints, but their LoS or lower-cadence nature leaves ambiguities in interpretation. For example, $B_l$ generally contains a mixture of horizontal and radial field components, so the changes of $B_h$ and $B_r$ generally cannot be distinguished. Furthermore, the default HMI vector magnetograms have a cadence of 12 minutes and a wider, tapered temporal averaging window of $\sim$22.5 minutes, so a large population of the rapid changes is not temporally resolved.

Another concern is the ``magnetic transients'', where the field variations are impulsive and temporary, contrary to the more permanent magnetic imprints \citep[e.g.,][]{kosovichev2001}. The changes, which sometimes appear as a brief $B_l$ sign reversal, seem to correlate with white-light or hard X-ray flare emission and are thought to be artifacts induced by flare-altered line profiles \citep{qiu2003,abramenko2004}. For the 12-minute-cadence HMI vector data processing, potentially anomalous line profiles may be averaged with normal ones, making diagnostics difficult.

To definitively characterize the rapid, vector field evolution, we have created a new high-cadence (90~s or 135~s) vector magnetogram dataset from HMI and use it to examine an archetypical event. Our intentions are twofold. Firstly, we provide a reference for the dataset by describing the key processing procedures and new features. Secondly, we demonstrate that the dataset is well suited for studying the magnetic imprints and transients in a quantitative and more temporally resolved manner. The new observations reveal a highly structured pattern of field evolution, which sheds light on the momentum processes in solar eruptions. We discuss the potential usage of the dataset for other studies.


\section{Data}
\label{sec:data}

HMI measures the Stokes parameters at six wavelengths in the photospheric \ion{Fe}{1} 6173~{\AA} absorption line. One of its two cameras is dedicated to the vector magnetic field. Under the original ``Mod-C'' observing scheme, vector magnetograms are inferred exclusively from this camera, and a full set of Stokes parameters $(I, Q, U, V)$ requires 135~s to complete. Since April 2016, HMI has been operating under a new ``Mod-L'' observing scheme, which combines the polarization measurements from both cameras (\citealt{liuy2016}; HMI Science Nugget \href{http://hmi.stanford.edu/hminuggets/?p=1596}{\#56}). This results in reduced noise in the LoS component and a shorter, 90~s observing cycle. Under both observing schemes, multiple sets of Stokes parameters are temporally averaged to suppress photon noise, leading to a routine vector magnetogram dataset with 720~s cadence.

We take advantage of the high-cadence Stokes measurement and create a new, full-disk vector magnetogram dataset with 135~s cadence (\href{http://jsoc.stanford.edu/ajax/lookdata.html?ds=hmi.B_135s}{\texttt{hmi.B\_135s}}). The 90~s cadence version is under development. The dataset has the same format as the standard 720~s version \citep{hoeksema2014} and is processed with identical pipeline options except the following.

\begin{enumerate}[parsep=0ex,partopsep=-0.5ex,itemsep=0.5ex,leftmargin=3mm]

\item The filtergrams are interpolated linearly in time, and all contributing filtergrams are taken within 270~s (``quick-look'' averaging scheme). This contrasts with the default, higher-order interpolation scheme and a wider temporal window that can produce artifacts when features are fast evolving \citep[e.g.,][]{oliveros2011}.

\item A 50~G constant is added to the noise masks that are used as weak-field threshold in the 180$^\circ$ azimuth ambiguity resolution algorithm \citep{hoeksema2014} to account for the higher noise (see below).

\item Data will be processed for selective periods of significant activity and by request only. The first release of $\sim$290~hr data covers about 30 events, most of which feature M- or X-class flares\footnote{For available time intervals and more details on the dataset, see \url{http://jsoc.stanford.edu/data/hmi/highcad}.}. The corresponding Stokes parameter dataset (\href{http://jsoc.stanford.edu/ajax/lookdata.html?ds=hmi.S_135s}{\texttt{hmi.S\_135s}}) is also available.

\end{enumerate}

The high-cadence data have higher noise due to shorter integration time and are more susceptible to contamination by $p$-mode oscillation. We illustrate this by comparing 135~s and 720~s data for AR 11158 at one instance (Figure~\ref{f:compare}(a)). The distribution of field strength $B$ in the 135~s data ($B_{135}$) peaks at 129~G, while the 720~s data ($B_{720}$) peaks at 85~G (Figure~\ref{f:compare}(b)). These are typical values in the quiet Sun where the polarization degree is low, and the inferred $B$ largely originates from photon noise. For $B_l$, the two frames are well correlated down to the deca-Gauss range (Figure~\ref{f:compare}(d)), suggesting that most noise resides in the transverse component. 

We have carried out a similar comparison for 257 pairs of 135~s and 720~s full-disk $B$ image over 6.4 days in February 2011. The median of $B_{135}$ varies daily between about 155 and 175~G, presumably induced by \textit{SDO}'s orbital velocity \citep{hoeksema2014}. It varies in phase with the median of $B_{720}$, and is consistently $\sim$50~G higher. We thus add 50~G to our noise mask for azimuth disambiguation.

In the example frame, the 135~s and 720~s data agree well in the strong-field regions. For pixels with $B>300$~G, the differenced $B$ ($\Delta B$) and $B_l$ ($\Delta B_l$) have narrow distributions centered around 0 (Figures~\ref{f:compare}(c)~and~(e)). The half width half maximum (HWHM) is 25 and 17~G for $\Delta B$ and $\Delta B_l$, respectively. For comparison, the median formal uncertainty of field strength $\sigma_B$ derived from spectral line inversion is 35 and 28~G for the 135~s and 720~s data, respectively. Evolution also contributes to the difference.

In this study, we focus on a 2 hr interval around an X-class flare on 2011 February 15, during which 54 frames of 135~s cadence data are available. We keep the images in the native Helioprojective-Cartesian coordinate, re-project the field vectors into a Heliocentric-spherical coordinate, and propagate the formal uncertainties \citep{sun2013}. We co-align the frames by cross-correlating continuum images obtained from inversion. The final dataset consists of cubes with a $250\arcsec\times170\arcsec$ field of view at a $0.\arcsec5$ pixel scale. 


\begin{figure*}[th!]
\centerline{\includegraphics{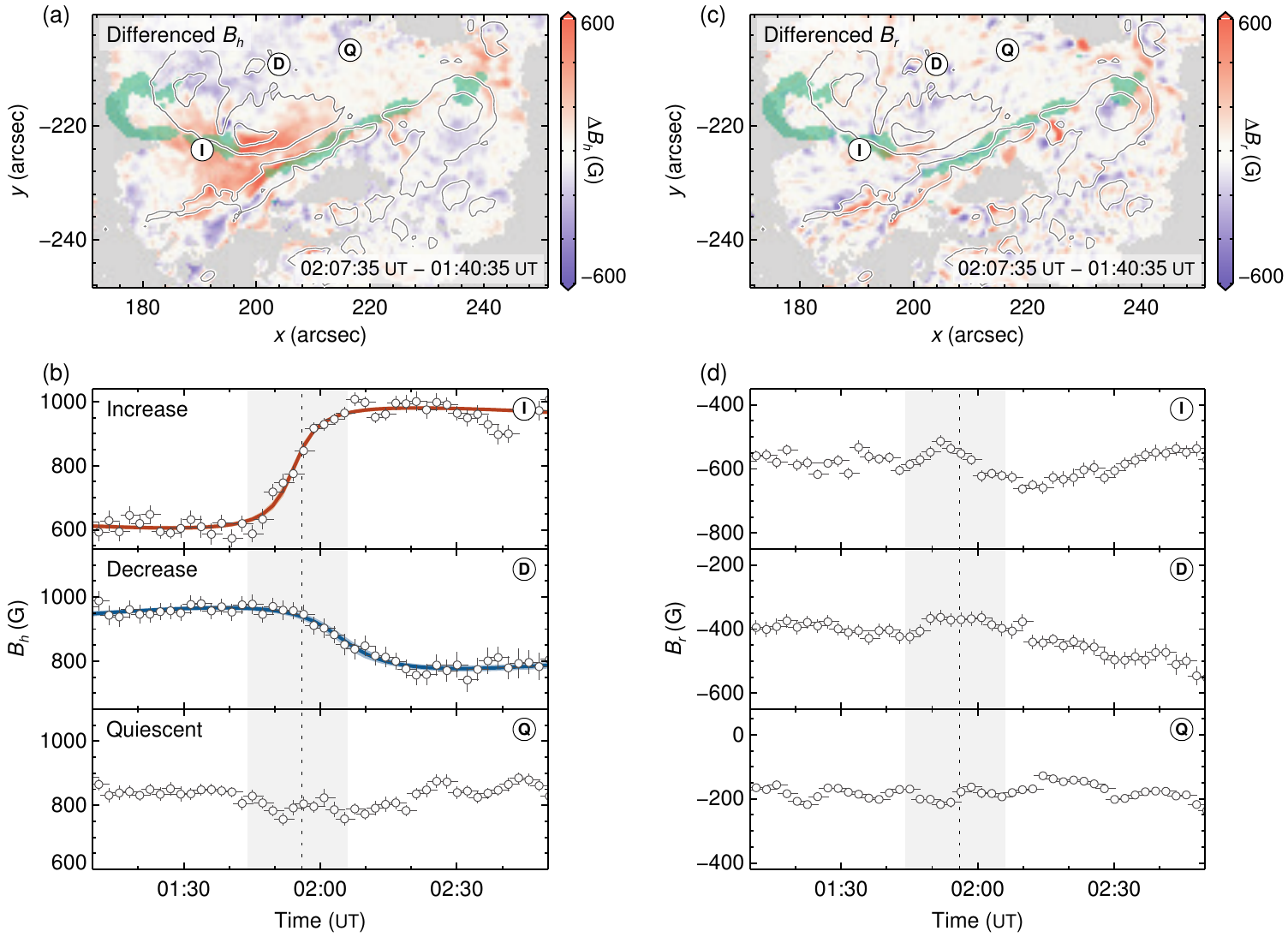}}
\caption{Rapid magnetic field evolution. (a) Base-differenced $B_h$ map. Gray regions have $B<300$~G. Green shades outline flare ribbons in AIA 1600~{\AA} at 01:47:54. Contours are for $B_r$ at $\pm1000$ and $\pm2000$~G. Symbols ``I'', ``D'', and ``Q'' mark the sample pixels, where $B_h$ exhibits significant increase, decrease, or little change (quiescent), respectively. (b) Temporal evolution of $B_h$ at sample pixels. Here and after, symbols show observations; horizontal error bars show the temporal averaging window of filtergrams; vertical error bars show formal uncertainty from spectral line inversion. Curves show a fitted step-like function. Vertical gray band indicates \textit{GOES} flare time; vertical dotted line indicates SXR flare peak. (c)--(d) Similar to (a)--(b), but for $B_r$. (An \href{http://vimeo.com/209908267}{animation} of this figure is available.)}
\label{f:profile}
\end{figure*}


\section{Results}
\label{sec:results}

AR 11158 generated the first X-class flare of Cycle 24, \texttt{SOL2011-02-15T01:56}. An X2 flare and a fast CME originated from the central bipole in this quadrupolar AR, located at W20S10 \citep{schrijver2011,sun2012}. The flare ribbons exhibited an archetypical ``double-J'' morphology (Figure~\ref{f:profile}(a)), which then extended both along and away from the main PIL. The \textit{GOES} SXR flare started, peaked, and ended at 01:44, 01:56, and 02:06, respectively. The \textit{RHESSI} 25--50~keV hard X-ray (HXR) flux peaked at 01:54, two minutes before the SXR peak. The magnetic imprints and transients have been studied using routine HMI 45~s LoS and 720~s vector data \citep[e.g.,][]{kosovichev2011,wangs2012a,sun2012,gosain2012,maurya2012,petrie2013,bayanna2014}. 

We now re-examine the more rapid magnetic evolution using the 135~s vector dataset. We focus on the scalers $B_h$ and $B_r$, and defer analysis of other variables such as azimuth and electric current to future studies. We utilize a new database for flare ribbons \citep{kazachenko2017} observed in 1600~{\AA} by the Atmospheric Imaging Assembly (AIA). It corrects for spurious intensities associated with strong flare emission and provides easy access to the evolving ribbon morphology.

\subsection{Example of Magnetic Imprint}
\label{subsec:example}

The co-aligned, high-cadence data now allow us to perform meaningful temporal analysis on single pixels. Following \citet{sudol2005}, we use a step-like function to model the magnetic imprint in a time sequence of field component $B_i$,
\begin{equation}
B_i(t) = a + b t + c \left\{ 1 + \frac{2}{\pi} \arctan [n (t - t_m)] \right\},
\label{eq:ebtel_n}
\end{equation}
where $a$, $b$, $c$, $n$, and $t_m$ are free parameters. The term $a + b t$ accounts for linear evolution; $\Delta B_i=2c$ measures the magnetic field change; $t_m$ corresponds to the mid-time of change; $\tau = \pi n^{-1}$ characterizes the time scale of change; $t_s=t_m-\tau/2$ is the start time of change; and $dB_i/dt=\Delta B_i / \tau$ is the change rate. We employ a least-square Monte Carlo method for fitting, and quote the median and 1$\sigma$ confidence interval when needed. To reduce the effect of noise, we consider only strong-field pixels, where $B>300$~G. Additional details of modeling are presented in Section~\ref{subsec:stats}.

A base-difference map of $B_h$ (Figure~\ref{f:profile}(a)) shows clear, structured patterns of field change. In particular, $B_h$ increases significantly near the main PIL between the flare ribbons, in agreement with previous findings \citep{wangs2012a,sun2012,gosain2012,petrie2013}. An equally important aspect is the wide-spread, though somewhat weaker decrease of $B_h$ further away from the PIL. A closer look at the temporal evolution of two representative pixels reveals clear, step-like changes that are well-resolved temporally (Figure~\ref{f:profile}(b)). The fit parameters $(\Delta B_h, t_m,\tau)$ are $(468_{-33}^{+35}, 10.2_{-0.5}^{+0.5}, 13.4_{-2.3}^{+2.5})$ and $(-330_{-103}^{+75}, 20.5_{-1.5}^{+1.6}, 28.8_{-10.5}^{+12.9})$ for the two example pixels (in units of G, minute since flare start, and minute), respectively. The increase is stronger, occurs earlier, and evolves faster compared to the decrease. In contrast, the example quiescent profile is not well fitted by the step-like function.

To assess the significance of $\Delta B_h$, we evaluate the secular evolution by differencing pairs of $B_h$ maps both before or both after the flare. We choose a time lag of 11.25~minute (5 frames), which is close to the median of $\tau$ (see Section~\ref{subsec:stats}), i.e., typical magnetic imprint time scale. For 15 pre-flare and 19 post-flare pairs, the root mean square (rms) $\Delta B_h$ is 64~G, and the rms formal uncertainty of $\Delta B_h$ is 68~G. We take the quadrature sum 93~G as the quiescent background. The changes at the two example pixels are thus at $5.0_{-0.4}^{+0.4}\sigma$ and $-3.5_{-1.1}^{+0.8}\sigma$, respectively.

In comparison, the variation of $B_r$ is less pronounced and less structured (Figure~\ref{f:profile}(c)). There appear to be patchy changes along the flare ribbons of both increase and decrease. The quiescent background of $\Delta B_r$ is 76~G. Neither example pixels with step-like $B_h$ change exhibit significant $B_r$ change (Figure~\ref{f:profile}(d)).%


\begin{figure*}[!th]
\centerline{\includegraphics{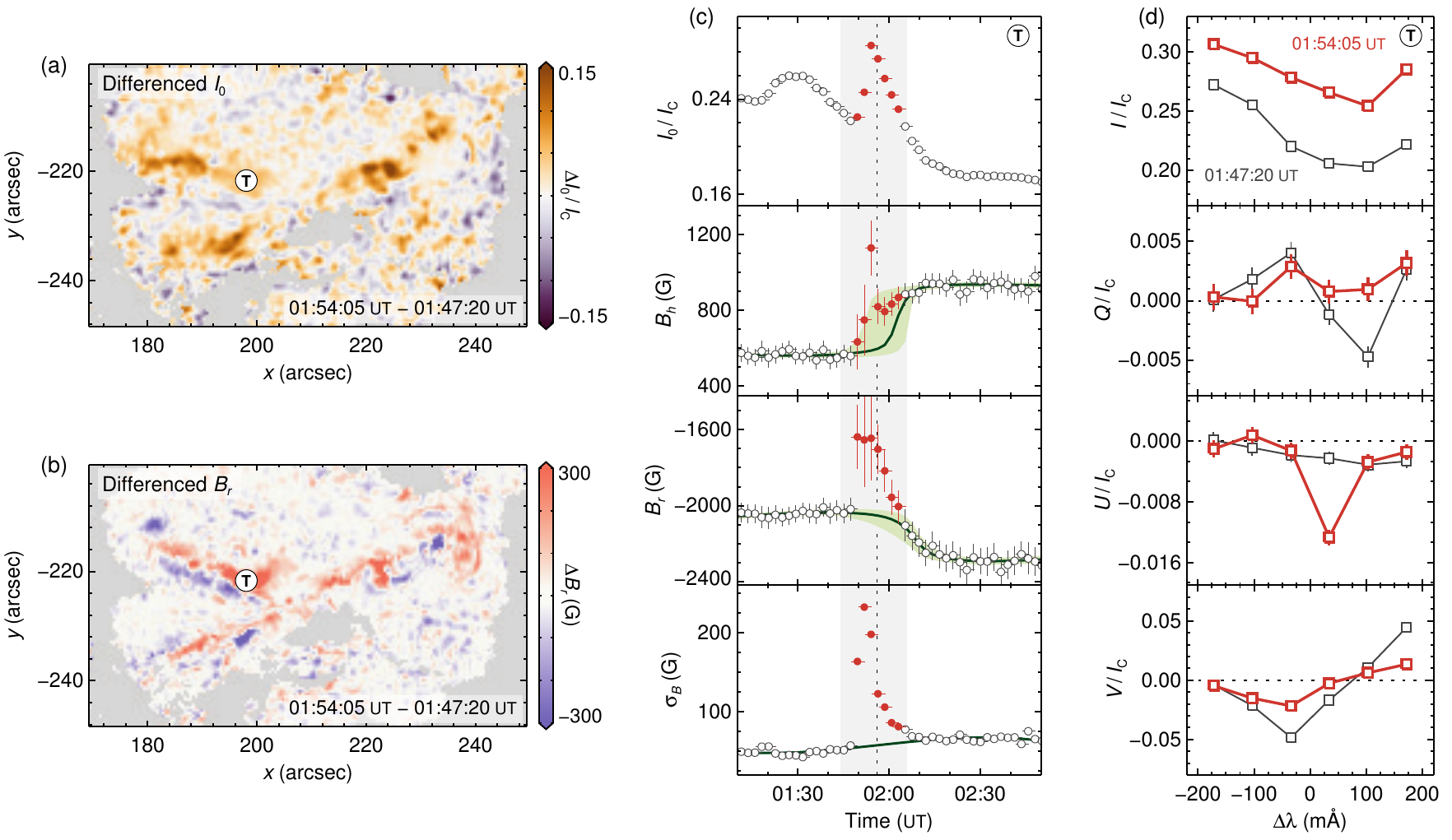}}
\caption{Flare-induced artifact as ``magnetic transient''. (a) Differenced map of Stokes $I_0$ ($\Delta \lambda = +172$~m{\AA} offset from line center) near the flare peak, normalized to the quiet-Sun mean continuum $I_c$. Symbol ``T'' marks the sample pixel. (b) Differenced $B_r$ map. (c) Temporal evolution of the sample pixel. Red symbols show the frames affected by flare emission. Green curves show the fitted step-like function for $B_h$ and $B_r$ and a fitted third-order polynomial for $\sigma_B$; green bands show the $1\sigma$ fitting confidence interval. Larger fitting uncertainty during flare time is due to the fact that we exclude magnetic transients. (d) Stokes profiles of the sample pixel at two instances, near (red) and before (gray) the flare peak. Error bars are derived assuming Poisson statistics.}
\label{f:artifact}
\end{figure*}


\subsection{Magnetic Transients}
\label{subsec:transients}

Magnetic transients have been reported for \texttt{SOL2011-02-15T01:56} using HMI 45~s LoS data. They are associated with continuum enhancement and Doppler-velocity transients \citep{kosovichev2011}. The observed left- and right-circular-polarization profiles appear to be distorted \citep{bayanna2014}. HMI high-cadence Stokes data have been used to study magnetic transients too. For this event, transient changes occur in all Stokes parameters \citep{maurya2012}. For an M7.9 flare \texttt{SOL2012-03-13T17:41}, transient changes in linear polarization appear to be consistent with genuine field evolution \citep{harker2013}. 

We search for transient signals first by inspecting running difference image sequences. In two elongated patches that resemble UV ribbons, Stokes $I$ increases across the line profile by as much as 15$\%$ of the nearby mean quiet-Sun continuum value ($I_c$) during the flare impulsive phase (Figure~\ref{f:artifact}(a)). Transient field changes appear in both $B_h$ and $B_r$ (Figure~\ref{f:artifact}(b)), and are approximately co-spatial with Stokes $I$ enhancement. We do not find obvious sign reversals in $B_r$.

Time profile of an example pixel (Figure~\ref{f:artifact}(c)) exhibits a resolved transient change in both $B_h$ and $B_r$ during the flare, superposed on a step-like, permanent change. This suggests that the magnetic transient can occur in conjunction with the magnetic imprint. The increase starts early in the flare and reaches maximum slightly before the SXR peak. We find that the formal uncertainty of inferred field strength $\sigma_B$ increases significantly during this period, nearly tripling the background. The Stokes profiles deviate from the pre-flare conditions (Figure~\ref{f:artifact}(d)); polarization generally becomes weaker except for $U$ near the line core. These observations suggest that the Stokes profiles are distorted by flare emission and are not adequately modeled under the default settings of the spectral line inversion algorithm \citep{centeno2014}. The inferred magnetic fields become less reliable.


\begin{figure*}[!th]
\centerline{\includegraphics{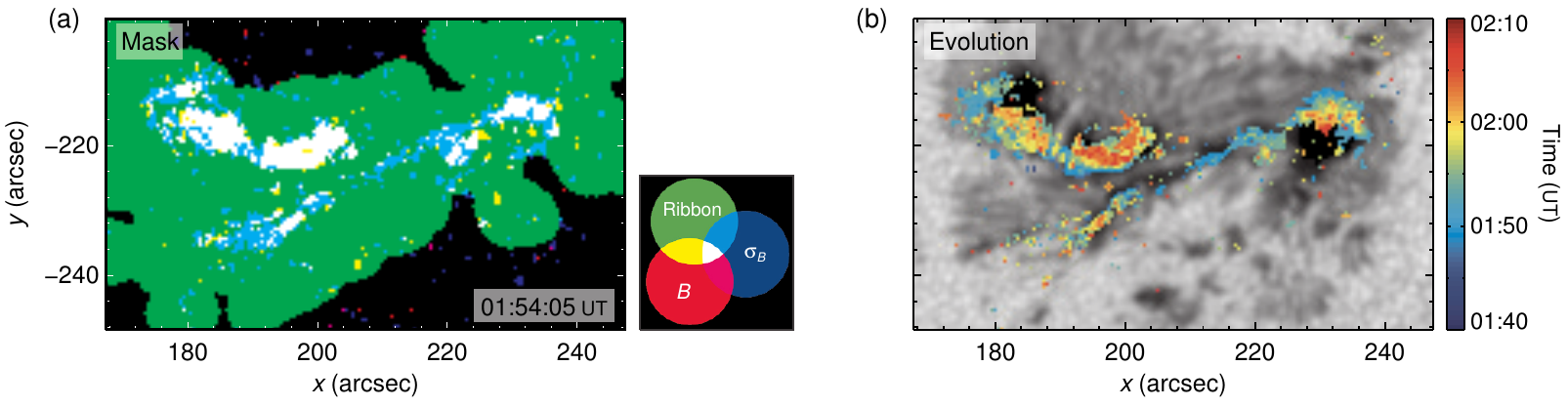}}
\caption{Identification and evolution of magnetic transients. (a) Masks for the 01:54:05 UT frame. AIA 1600~{\AA} instantaneous flare ribbon mask is shown in green; mask for transient values in $B_r$ or $B_h$ time sequence is shown in red; mask for transient values in $\sigma_B$ is shown in blue. Pixels where all three masks overlap are colored white and used as our final mask. (b) Evolution of the transient mask color-coded by time, overplotted on a pre-eruption $I_0$ map. (An \href{http://vimeo.com/209908274}{animation} of this figure is available.)}
\label{f:mask}
\end{figure*}


To identify and characterize the magnetic transients, we apply the following four criteria with respective empirical identification schemes (Figures~\ref{f:artifact}(c) and~\ref{f:mask}(a)).

\begin{enumerate}[parsep=0ex,partopsep=-0.5ex,itemsep=0.5ex,leftmargin=3mm]

\item The observation is made during the \textit{GOES} flare time.

\item The pixel resides in the flare ribbons. The UV ribbons at each instant are expected to be much more extended than the white-light sources, and thus should safely encompass the impacted photospheric region. Our new AIA 1600~{\AA} flare ribbon database provides co-aligned masks of ribbon locations at a 24~s cadence \citep{kazachenko2017}, which we further dilate by $\sim$3~Mm (8 pixels). For each HMI measurement time $t$, we create a ``ribbon mask'' that includes all pixels in the AIA masks within $t\pm2$~minute.

\item The formal uncertainty of field strength, $\sigma_B$ significantly exceeds the non-flaring background in a single-pixel time sequence. We mask out the time steps during the flare and fit a third-order polynomial to the rest, assuming there is no sudden change in measurement quality during quiescent times. We mark those flare-time measurements that exceed the fit by more than three times the rms residual. For a single time step, all marked locations constitute a ``$\sigma_B$ mask''.

\item The measured magnetic field $B_h$ or $B_r$ is an outlier in a single-pixel time sequence. We mask out all measurements of suspect quality identified in the previous step, and fit the rest with both a step-like function and a third-order polynomial. Using the better of the two models (smaller reduced chi-square $\chi_r^2$), we mark those flare-time measurements that deviate from the fit by more than three times the rms residual, and create a ``$B$ mask'' for each time step.

\end{enumerate}

The $\sigma_B$ mask (blue in Figure~\ref{f:mask}(a)) and $B$ mask (red) reside almost completely inside the ribbon mask (green), suggesting that the transient variations of the field measurement and its quality are indeed correlated with UV flare emission. The $\sigma_B$ and $B$ masks often overlap. However, there are cases where the quality of inversion is suspect but no transient signal is found in the inferred magnetic field (cyan). There are also cases where the field changes transiently but the quality of inversion remains similar (yellow), so there is no evidence against genuine field evolution. A detailed analysis of the Stokes parameters and the inversion result at these locations is necessary, but is out of the scope of this work.

Here, we narrowly define magnetic transients as measurements that satisfy all four criteria above (white in Figure~\ref{f:mask}(a)). By definition, they appear where the magnetic field cannot be reliably derived from flare-impacted Stokes observations. They cover about 6$\%$ of strong-field pixels in AR 11158. Our empirical scheme appears to work effectively at separating transients from magnetic imprints and secular evolution (Figure~\ref{f:artifact}(c)). Interestingly, the location and evolution of the identified transients (Figure~\ref{f:mask}(b)) closely resemble that of the white-light sources in \textit{Hinode} continuum observations \citep{kerr2014}, even though we have not explicitly used HMI Stokes $I$ or continuum in our scheme. This further suggests that the photospheric impact of flare emission is a necessary condition for magnetic transients.

Transients identified in HMI LoS observations are similar in nature, as the essential assumption of a Gaussian line profile may break down. Additional artifacts may also come from our observing scheme. For example, the slightly different observation times of the Stokes parameters at different wavelengths can cause an undesirable aliasing effect \citep{oliveros2014}.


\begin{figure*}[!th]
\centerline{\includegraphics{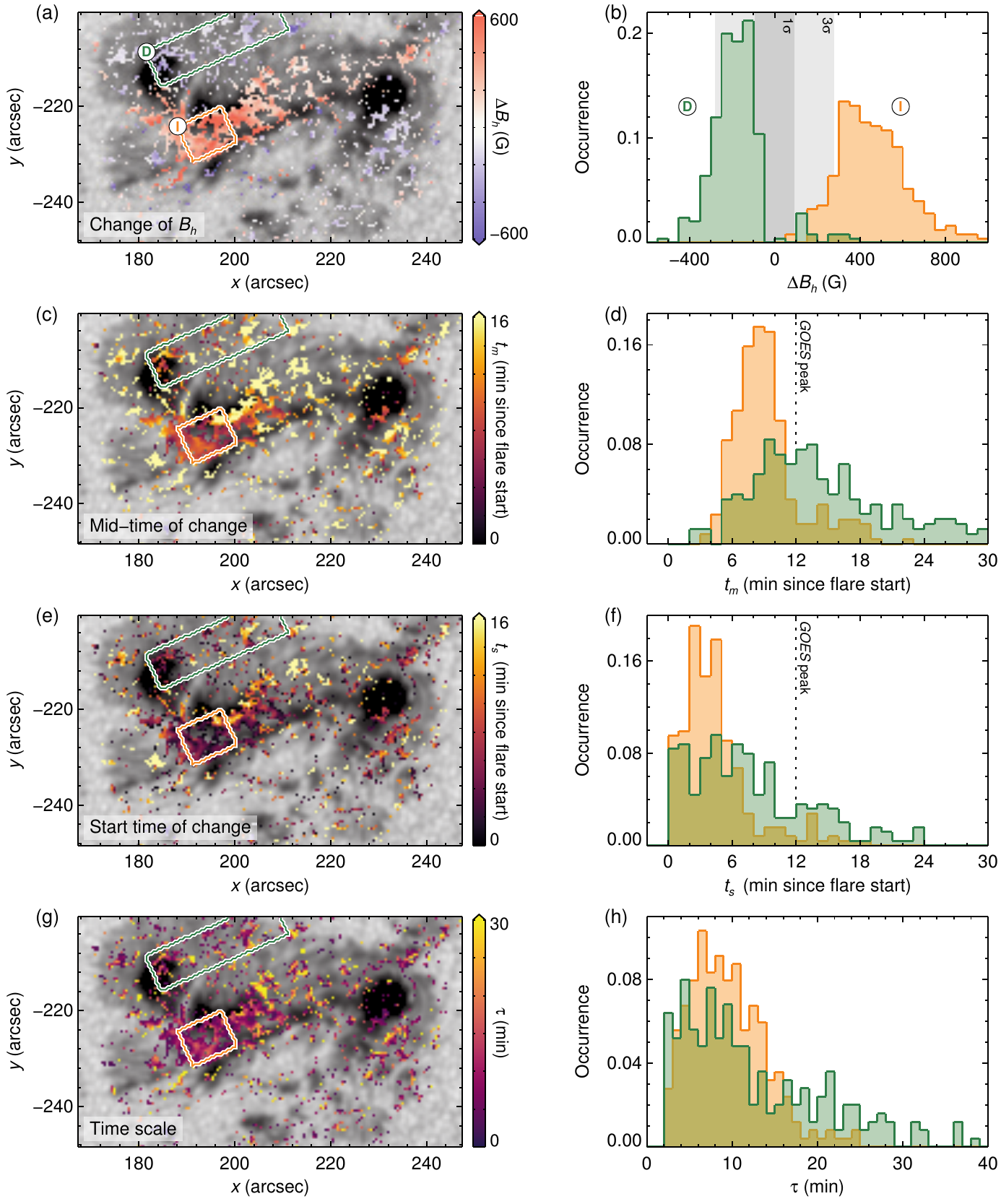}}
\caption{Characteristics of the field evolution derived from the step-like function fit after correcting for magnetic transients. (a) Horizontal field change $\Delta B_h$, overlaid on a pre-eruption $I_0$ map. Only pixels with reasonable fits are included (see the text for details). The two subregions marked as ``I'' and ``D'' are selected for comparison. (b) Histogram of $\Delta B_h$ in subregions ``I'' and ``D''. Darker and lighter gray bands indicate $1\sigma$ and $3\sigma$ quiescent background, respectively. (c)--(d) Similar to (a)--(b), but for the mid-time of change $t_m$. Vertical dotted lines in (d) indicates \textit{GOES} SXR peak time. (e)--(f) Similar to (a)--(b), but for the start time of change $t_s$. (g)--(h) Similar to (a)--(b), but for the time scale of change $\tau$.}
\label{f:stats}
\end{figure*}


\subsection{Statistics of Magnetic Imprints}
\label{subsec:stats}

We now study the statistical behavior of magnetic imprints. After excluding the identified transient measurements, we fit the time sequence at each pixel with both a step-like function (for magnetic imprints) and a third-order polynomial (for secular evolution). Only pixels that favor the magnetic imprint model, that is, having a smaller $\chi_r^2$ for the step-like function fit, are included in our analysis. About 5$\%$ of pixels with the poorest fit ($\chi_r^2\ge5.5$) are discarded.

We further apply several empirical selection criteria. To ensure that the profile is temporally resolved, we include only pixels where the time scale is longer than the cadence ($\tau\ge135$~s) and the mid-change time is no earlier than the first observation since flare start ($t_m\ge1.08$~minute)\footnote{In practice, fitting is performed within the following limits to ensure a physically meaningful imprint model: $1.08 \le t_m \le 37.08$~minute (from 01:45:05 to 02:21:05) and $2.25 \le \tau \le 63.62$~minute (from 1 to 9$\pi$ time steps). Fits hitting any limit (e.g., $t_m=1.08$ or $\tau=2.25$) are excluded.}. We additionally require that the change starts after the flare onset ($t_s\ge 0$), but no too long after ($t_m<36~$minute and $\tau<1~$hr). The rate of field change of the magnetic imprint should also exceed that of the linear evolution $|dB/dt|>|b|$ (see Equation~\ref{eq:ebtel_n}).

For $B_h$, 15$\%$ (about 4200) strong-field pixels are finally selected (Figure~\ref{f:stats}). In particular, we compare two subregions (Box ``I'' and ``D''), which contain about 250 well fitted pixels each. Our analysis indicates the following. 


\begin{figure*}[!th]
\centerline{\includegraphics{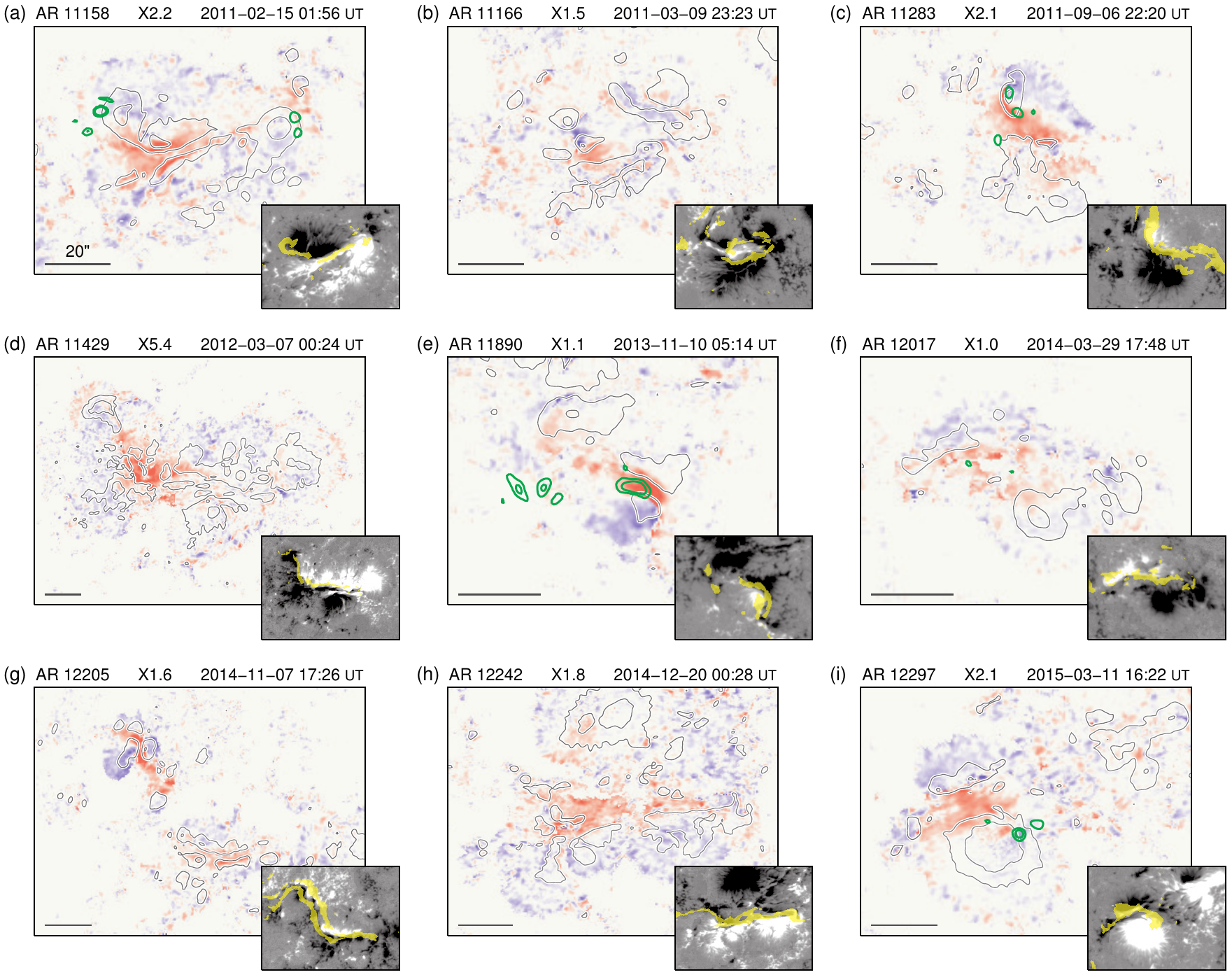}}
\caption{Overview of magnetic imprints in nine arbitrarily selected events with X-class flares from the first data release. Each panel shows base-differenced $B_h$ map scaled between $\pm$600~G with $B_r$ contours at $\pm$1000 and $\pm$2000~G, same as Figure~\ref{f:profile}(a). Insets show $B_r$ maps scaled between $\pm$1000~G with 1600~{\AA} flare ribbons observed early during the flare overlaid. In panels (a), (c), (e), (f), and (i), green contours show the 6~mHz egression power at 3$\sigma$ and 6$\sigma$ (quiet-Sun rms) around flare peak as possible sunquake sources \citep{chen2016}. Scale bars are 20{\arcsec} in all panels.}
\label{f:collage}
\end{figure*}


\begin{enumerate}[parsep=0ex,partopsep=-0.5ex,itemsep=0.5ex,leftmargin=3mm]

\item Magnetic imprints appear over the entire AR. Most imprinted pixels are located in the inner or the outer penumbra of the central sunspot pair. The former resides along the PIL between the flare ribbons; the latter brackets the ribbons.

\item The magnitude of change $\Delta B_h$ is almost exclusively positive in Box ``I'' and negative in Box ``D'' (Figures~\ref{f:stats}(a) and (b)). Box ``I'' has a median increase of 441~G ($4.7\sigma$, the quiescent background). Box ``D'' shows a weaker decrease with a median of $-174$~G ($1.9\sigma$). While $\Delta B_h$ may not be significant at individual pixels, the wide-spread, coherent pattern of change is striking. A two-sample Kolmogorov-Smirnov (K-S) test\footnote{We compare the distribution of $\Delta B_h$ in a difference $B_h$ map spanning the eruption (Figure~\ref{f:profile}(a)) with seven difference maps before the eruption. The K-S test median $K$ is $0.510$, and in all cases $p \ll 10^{-5}$. We thus reject the null hypothesis that $\Delta B_h$ of magnetic imprints and quiescent evolution are drawn from the same distribution.} on $\Delta B_h$ in Box ``D'' confirms that the changes during the eruption are very different from quiescent evolution.

\item The increases in Box ``I'' occur early during the flare, with median $t_m$ and $t_s$ of 8.6 and 3.8~minutes (since flare start; Figures~\ref{f:stats}(c)--(f)), respectively. These are much earlier than the SXR peak at 12~minutes and the HXR peak at 10~minutes. The decreases in Box ``D'' occur slightly later. Parameters $t_m$ and $t_s$ have a wider distribution and median of 13.1 and 6.5~minute, respectively. Almost all pixels (99$\%$) start changing during the flare ($t_s<22$~minute). 

\item The median time scale of change $\tau$ is 8.9~minute for Box ``I'' and 10.1~minute for Box ``D'' (Figures~\ref{f:stats}(g) and (h)). The decreases occur more slowly, as about 22$\%$ of pixels have $\tau>20$~minute.

\end{enumerate}

These results and our selection criteria deserve some discussion. Firstly, the magnetic imprints appear to be spatially separated from the magnetic transients despite some overlap (see Figures~\ref{f:stats}(a) and~\ref{f:mask}(b)). About 40$\%$ of transient pixels are co-spatial with imprints, while only 9$\%$ of imprints are marked for transients. Secondly, our requirement that the field change must occur after the flare onset is purely empirical, which aims to establish some causal relations between the imprint and the flare. However, many excluded pixels (28$\%$ of the final selection) satisfy all other criteria but have $-5 \le t_s < 0$~minute. Given the measurement uncertainties, it is possible that they are genuine magnetic imprints. It is also possible that magnetic evolution can indeed precede flare onset. Thirdly, a significant number of excluded pixels (89$\%$ of the final selection) have a small $\tau$; their step-like changes are not resolved at HMI's cadence. This is compatible to $B_l$ observations, where $\sim$50$\%$ \citep{sudol2005} and $\sim$25$\%$ \citep{petrie2010} of all events occur on a time scale of less than 2 minutes. Fourthly, the spatial distribution of $t_m$ (Figure~\ref{f:stats}(c)) suggests that the $B_h$ changes ``propagate'' across the AR from the main PIL, similar to the findings in \cite{sudol2005} using $B_l$ observations.

We apply the same procedures on $B_r$ and find many pixels with clear step-wise changes significantly above the quiescent background (for a marked example, see Figure~\ref{f:artifact}(c)). Nevertheless, the changes appear much less structured spatially and temporally. We do not attempt to make further conclusions.


\section{Discussion and Outlook}
\label{sec:discussion}

The new high-cadence vector dataset allows us to quantitatively depict a scenario where flare-associated magnetic imprints, mainly appearing as step-wise, persistent changes in $B_h$, occur over the entire AR with a spatially and temporally structured pattern. Along the main PIL, $B_h$ increases rapidly during the early phase of the flare, whereas in the AR periphery $B_h$ decreases more slowly and at later times. The field change is typically a few hundred Gauss, well above the quiescent background evolution, and stronger for the increase than decrease. The time scale of temporally resolved changes is about 10 minutes; a significant portion is still unresolved at HMI's 135~s cadence.

We note that detailed temporal analysis has hitherto been limited to $B_l$. Depending on the AR's location, $B_l$ can include large contributions from the less varying $B_r$, so the pattern of field change may not be obvious. Moreover, although the contrasting behaviors of $B_h$ in the AR core and the periphery were previously noticed in differenced vector data \citep{wangjx2009}, the crucial temporal information was missing.

Our new dataset is capable of removing certain ambiguities arising from LoS only or lower-cadence vector observations. For example, \cite{petrie2010} suggested that the observed step-wise $B_l$ changes mainly result from the horizontal field changes based on the fact that the LoS flux varies more when the AR is closer to the limb. HMI 720~s vector data support the claim \citep{wangs2012b,petrie2012}, but lack the temporal information to reproduce the step-shaped profiles seen in the 1-minute-cadence $B_l$ data. This can now be verified by decomposing the 135~s field vectors and comparing the more temporally resolved behaviors of $B_l$ and $B_h$. The high cadence and the information returned from spectral line inversion also allow us to effectively separate magnetic imprints from transient signals. We can thus comment on the genuineness of the flare-related field changes.

Is the picture above universal? Preliminary inspection of nine ARs hosting X-class flares suggests a positive answer (Figure~\ref{f:collage}). Many other aspects of magnetic imprints beside $B_h$ are worth exploring too. Are imprint characteristics correlated to the flare \citep{wangs2012b} and CME properties \citep{sun2015}? How do the azimuth \citep{petrie2013,harker2013}, electric current \citep{janvier2014}, and magnetic topology \citep{zhaoj2014} evolve? Follow-up surveys are straightforward and are poised to address these questions. 

New advances on magnetic imprint and transient study may come from high-spectral-resolution observations or more sophisticated magnetic field inference techniques. \cite{kleint2017} reported step-wise changes in chromospheric $B_l$ for an X1 flare (\texttt{SOL2014-03-29T17:48}) using DST/IBIS \ion{Ca}{2} 8542~{\AA} observations. The changes appear uncorrelated to their photospheric counterparts in HMI $B_l$ (for $\Delta B_h$, see Figure~\ref{f:collage}(f)). \cite{kuckein2015} studied the photospheric and chromospheric responses in an M3 flare (\texttt{SOL2013-05-17T08:57}) using \ion{Si}{1} 10827~{\AA} and \ion{He}{1} 10830~{\AA} triplet observed with VTT/TIP-II. Full inversion of the \ion{Si}{1} Stokes shows that the field strength decreases temporarily during the flare but recovers afterwards. These results illustrate a more complicated picture than that proposed above, which warrants further investigation. Upcoming NST and DKIST telescope magnetic field observations will contribute to this topic.

The origin of the magnetic imprints is not entirely clear. The coronal implosion conjecture \citep{hudson2000} is often cited to explain the increase in horizontal photospheric field. We note that the model mainly concerns the contracting coronal structure; it is not guaranteed that the photosphere responds in a similar fashion. As mentioned above, even the chromospheric and photospheric field evolution seems to be dissociated \citep{kleint2017}. Numerical models that reproduce the implosion phenomenon may help address the issue \citep{zuccarello2017}.

Below, we discuss a couple of implications from our results. The first point is also an attempt to explain the observation in terms of momentum conservation.

Firstly, we note that the total Lorentz force inside a volume can be expressed as a surface integral of the Maxwell stress tensor on its boundaries, fully determined by the local magnetic field \citep{fisher2012}. If we choose a volume in the solar atmosphere that encloses the entire CME ejecta, place its lower boundary in the photosphere, and assume that the contribution from the side and top boundaries is negligible or largely invariant, the impulsive Lorentz force thought to provide the upward momentum of a CME \emph{must} manifest as the photospheric field changes. The increases of $B_h$ near the PIL will lead to a positive increase of the total vertical force $F_r \propto \sum (B_h^2 - B_r^2)$, which presumably drives the ejecta. It should be canceled later by a decrease of $B_h$ in the periphery if the volume is to return to force equilibrium. In other words, the observed rapid magnetic imprint that evolves on a coronal Alfv\'{e}nic time scale is a natural consequence of momentum conservation. In reality, gravitational force and thermal dynamics responses of the dense lower atmosphere complicate the situation \citep{sun2016}. We note that this putative upward Lorentz force \emph{inside} the volume should not be confused with the downward force exerted on the rest of the Sun by the selected volume. The latter is thought to be one possible mechanism for sunquakes (see below).

Secondly, numerical simulations of solar eruptions can be used to verify the arguments above. We have investigated the magnetic field evolution in the lowest layers of two published MHD models \citep{torok2005,lynch2009}. Preliminary analysis \citep{sun2016,lynch2017} shows that both display clear magnetic imprints similar to that of AR 11158, that is, $B_h$ increases in the AR core and decreases in the periphery, despite very different magnetic topology and eruption mechanisms. An earlier study of a third MHD model \citep{fan2010} showed similar signatures \citep{liyx2011}. None of these three models make assumptions that are known to produce magnetic imprints, and the agreement is unlikely a mere coincidence. We thus conjecture that the magnetic imprint may be a fundamental aspect of solar eruption.

We finally point out that the high-cadence vector magnetograms can be useful to the study of sunquakes and data-driven modeling of the solar corona, among other topics.

Sunquakes, a helioseismic response to the flare impact in the solar photosphere, have been thought to originate from high-energy electrons, protons, or radiative back-warming \citep[e.g.,][]{kosovichev1998,donea2005,zharkova2007}. Magnetic force was recently proposed as an alternative mechanism \citep{hudson2008,fisher2012}. The new explanation is particularly appealing for the sunquake observed in AR 11158, because the disturbance is observed before significant HXR emission, thus disfavoring a high-energy particle origin \citep{kosovichev2011}, and the sources appear to be co-spatial with two ends of the erupting flux rope \citep[Figure~\ref{f:collage}(a);][]{zharkov2011}. Nevertheless, studies of individual events have not reached a consensus \citep[e.g.,][]{alvaradogomez2012,judge2014}. To this end, a survey of sunquakes in the context of magnetic field variations will be helpful. A preliminary analysis \citep{chen2016} detects sunquake signals in five of the nine X-class flares illustrated here (Figure~\ref{f:collage}). The location, strength, and timing of the sources can now be compared with the magnetic evolution. Predictions from theoretical and numerical studies \citep[e.g.,][]{lindsey2014,russel2016} regarding the role of specific magnetic configuration can also be tested.

Knowledge of the coronal magnetic field is vital to our understanding of solar eruptions and our capability to predict major space weather events. New-generation data-driven models \citep[e.g.,][]{cheung2012,inoue2014,fisher2015,galsgaard2015,jiangcw2016} aim to take advantage of the observed evolution of the magnetic and velocity fields and model the evolution of the coronal field with sufficient accuracy and efficiency. \cite{leake2017} have investigated the effect of the driving time scale, i.e., the input data cadence, on the modeling accuracy using their newly developed, data-driven MHD framework. They drive the new model with photospheric conditions sampled from a ``ground-truth'' flux-emergence MHD simulation \citep[e.g.,][]{leake2013} and compare the outcomes with the known ground-truth. Rapid evolution of the sub-AR magnetic field cannot be recreated from a 12-minute-cadence driver. Contrarily, a 1.2-minute-cadence driver reduces the relative error in magnetic free energy by almost two orders of magnitude, down to less than 10$\%$. The test demonstrates that the high-cadence vector data are more suited for data-driven modeling, although the higher noise can be a concern.




\acknowledgments
We thank S\'{e}bastien Couvidat, Rebecca Centeno, Monica Bobra, Jeneen Sommers, and Hao Thai for assistance with data processing. This work is supported by NASA contract NAS5-02139 (HMI), NASA awards NNX13AK39G (CGEM) and NNH14ZDA001N-HGI, and NSF SHINE award AGS1622495. The \textit{SDO} data are courtesy of NASA and the \textit{SDO}/HMI science team. Function fitting is performed with MPFIT (\url{http://purl.com/net/mpfit}).

\facility{\textit{SDO}}


\end{CJK}




\end{document}